# Multi-pulse Photoacoustic Microscopy Based on k-space Pseudospectral Method: Simulation Study

Xianlin Song [a, #, *], Jianshuang Wei [b, c, #], Jianmin Xiong [a], Lingfang Song [d]

[a] School of Information Engineering, Nanchang University, Nanchang, China;

[b] Britton Chance Center for Biomedical Photonics, Wuhan National Laboratory for Optoelectronics-Huazhong University of Science and Technology, Wuhan, China;

[c] Moe Key Laboratory of Biomedical Photonics of Ministry of Education, Department of Biomedical Engineering, Huazhong University of Science and Technology, Wuhan 430074, China;

[d] Nanchang Normal University, Nanchang 330031, China;

[#] equally contributed to this work

*songxianlin@ncu.edu.cn

**Abstract**：Photoacoustic imaging has emerged in the past decades. Compared with the traditional medical imaging mode, it has better imaging performance and has great development potential in the field of biological imaging. In traditional photoacoustic microscopy, a single laser pulse is generally used to irradiate the sample to produce photoacoustic signal. And signal-to-noise ratio (SNR) is a very important indicator for photoacoustic imaging. In order to obtain the image with high SNR, multiple acquisition or increasing laser pulse energy is usually adopted. The former will lead to slower imaging speed, and the latter will lead to photobleaching or phototoxicity. Here, we proposes multi-pulse photoacoustic microscopy, the photoacoustic signals were stimulated sequentially using multiple laser pulses in each A line data acquisition. In order to verify the feasibility of this method, a multi-pulse photoacoustic imaging simulation platform is established using k-Wave toolbox. The performance of multi-pulse photoacoustic imaging is verified through the three scanning modes of photoacoustic microscopy A-scan, B -scan, and C-scan. The results indicate that the SNR is proportion to the number of laser pulses used, high SNR can be achieved by low-energy laser pulse. This work will help to expand the application of photoacoustic imaging.

## 1 Introduction

Medical imaging is of great significance to the diagnosis and treatment of illness. Experts have developed several medical imaging methods, including ultrasound imaging [1], X-ray imaging [2], and magnetic resonance imaging technology [3]. Ultrasound imaging technology uses the difference in absorption and reflection of human tissues with respect to ultrasonic waves, and receives and processes ultrasonic signals reflected by human tissues and organs to obtain images of internal organs. Compared with pure optical imaging, this imaging method has deeper penetration; X-ray imaging technology obtains the absorption characteristics of roentgen rays by different tissues by detecting the inten-sity of roentgen rays passing through biological tissues, and judges normal tissues and organs and lesions. However, this imaging method requires the use of a contrast agent to image the organs, which will endanger the health of the human body and has a certain degree of danger; the magnetic resonance technology is based on the development of the magnetic resonance phenomenon of the atomic nucleus, and the human body infor-mation is reconstructed through a computer. Due to the high cost of its equipment, it cannot meet people's needs.

Photoacoustic imaging is a new medical imaging method based on the photoacoustic effect [4]. The photoacoustic effect was proposed in 1880 by Bell. A laser pulse irradiates the sample, the sample absorbs its energy, the temperature rises instantaneously, and adiabatic expansion occurs, which then generates ultrasonic signal. Because the ultrasonic signal is excited by laser pulses, it is called photoacoustic signal. The photoacoustic effect can be successfully applied to biological tissue imaging because of the large differences in the absorption of laser pulses in different biological tissues [5] [6]. The photoacoustic signal received by the ultrasonic probe contains the absorption characteristics of laser pulses by biological tissues. The optical absorption distribution can be reconstructed by the reconstruction algorithm.

Photoacoustic imaging combines the advantages of pure optical and pure ultrasound imaging[7] [8]. Compared with pure optical imaging, it is no longer limited by light scattering and retains its advantages of high contrast; compared with pure ultrasound imaging, it retains its superior penetration performance and high imaging resolution. Compared with traditional medical imaging methods, photoacoustic imaging has outstanding advantages. But there are still many problems that need to be solved urgently. In photoacoustic imaging, the photoacoustic signal will inevitably be interfered by noise, so the signal received by the probe is the result of the mixture of photoacoustic signal and noise, the signal-to- noise ratio is not high, and the acquired deteriorating images have great interference in the diagnosis and treatment of diseases. It is very important to obtain highquality photoacoustic images.

In order to improve the signal-to-noise ratio of photoacoustic images, research teams around the world have developed a variety of methods. Yang et al. proposed a spatially invariant resolution photoacoustic microscope SIR-PAM without movement. It overcomes the limitation of 2D single-pixel Fourier spectrum acquisition and imaging without depth resolving ability, and retains the good characteristics of single-pixel imaging that can greatly improve the signal-to-noise ratio [9]. Image fusion methods are also used to improve photoacoustic image quality. Beckmann et al. used multi-wavelength superposition to improve the signal-to-noise ratio of photoacoustic images, and used a computer to extract the information obtained from different channels of the same image to the greatest extent, and then merged into one image. Compared with the original image, the quality of the image obtained by this method is higher, and the image contains more useful information [10].

In this manuscript, we proposes multi-pulse photoacoustic microscopy, the photoa-coustic signals were stimulated sequentially by multiple laser pulses in each A line data acquisition. In order to verify the feasibility of this method, a multi-pulse photoacoustic imaging simulation platform is established using k-wave toolbox. The performance of multi-pulse photoacoustic imaging is verified through the three scanning modes of photoacoustic microscopy A-scan, B -scan, and C-scan. The results indicate that the SNR is proportion to the number of laser pulses used, only multiple low-energy laser pulses are needed to obtain a high SNR.

**2 Simulation Platform of Multi-pulse Photoacoustic Microscopy**

**2.1 Pseudo-Spectral and k-Space Methods**

The k-Wave toolbox based on Pseudo-Spectral was released in 2009 by Bradley Treeby and Ben Cox [11]. It can solve coupled first-order wave equations in one, two, and three dimensions. The pseudo-spectral method to solve differential equations consists of two parts: one is discretization, and the other is calculating spatial derivatives. Establish a grid in the simulation area and use the interpolation method to get the value on the grid point; then, use the appropriate basis function to establish the interpolation function, and calculate the spatial derivative of the variables on the grid. It should be noted that the use of pseudospectral method is only limited to the spatial domain.

The k-Wave simulation includes kspaceFirstOrder1D, kspaceFirstOrder2D, and kspaceFirstOrder3D simulation functions, which are used to simulate acoustic wave simulation in one-dimensional, two-dimensional, and three-dimensional media, respectively. In the simulation, only four parameters need to be set: kgrid, medium, source and sensor, which define the characteristics of the computing grid, propagation medium, sound source, and ultrasonic detector, respectively. When these simulation functions are called, the propagation of sound pressure in the medium will be calculated step by step according to the time step size. Then the generated pressure value will be recorded by the ultrasonic transducer. The simulation ends when it runs to the specified number of time steps.

The simulated environment was created in two dimension with 448 × 448 pixels (each pixel size is 0.25 μm), and contains a perfectly matched boundary layer (PML) to satisfy the boundary conditions for the forward process. The samples are set as required. The number of ultrasonic transducer elements is 1, the center frequency is 50 MHz, and the bandwidth is 80%. It is placed 50.5 μm away from the center of the sample. The surrounding medium is water with a sound velocity of 1.5 km/s and a density of 1000 kg/m$^3$. All simulations assume that an acoustically homogeneous medium was considered with no absorption or dispersion of sound.

## 2.2 Multi-pulse photoacoustic imaging simulation model

We set up a multi-pulse model in the k-Wave toolbox, as shown in Figure 1. Using fiber delay method to divide a laser pulse into 6 laser pulses with a time interval of 300 ns each other. Noise was added during simulation. A-scan, B -scan, and C-scan were carried out to demonstrate the validity of the method. In A-scan, a ball with a radius of 2.5 μm is set in the center of the grid, the ultrasonic transducer is 50.5 μm away from the center of the grid to collect the photoacoustic signal of the ball, as shown in Figure 1. In B-scan, a carbon fiber is used as the sample, the laser pulse and the ultrasonic transducer are kept coaxial, and the B-scan image can be obtained by scanning point by point in steps. In C-scan, a blood vessel is used as the sample, the laser pulse and the ultrasonic transducer are kept coaxial, and the C-scan image can be obtained by 2D raster scan.

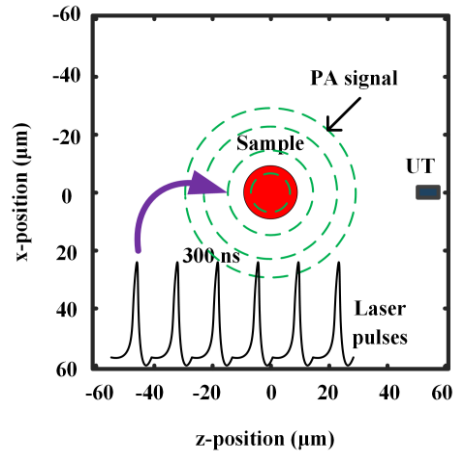

Figure 1. Schematic diagram of simulation platform of multi-pulse photoacoustic microscopy. UT, ultrasonic transducer.

## 3 Simulation results of multi-pulse photoacoustic imaging

### 3.1 A-Scan: performance of simulation platform

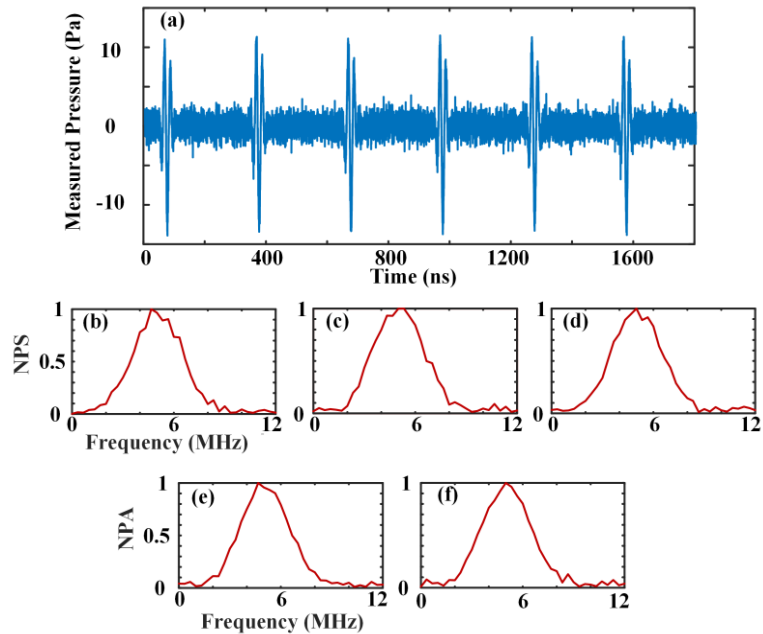

Figure 2. A-Scan: performance of simulation platform. (a) The photoacoustic signals excited by the 6 laser pulses are separated by 300 ns from each other. (b)-(f) Photoacoustic spectrum of the last five photoacoustic signals in (a). NPS, normalized power spectrum.

The photoacoutic signal in A data acquisition is shown in Figure 2. The photoacoustic signals excited by the 6 laser pulses are separated by 300 ns from each other, As shown in Figure 2(a). Extracting the last five photoacoustic signals, and performing spectrum analysis on them, as shown in Figures 2(b)-2(f). The distributions of the 5 photoacoustic signals in the frequency domain are the same, and the center frequency of the signal spectrum is 5MHz, the spectral range is 0-10 MHz, thus, these 5 photoacoustic signals can be superimposed on each other. Furthermore, it provides a basis for improving the signal-to-noise ratio through signal superposition.

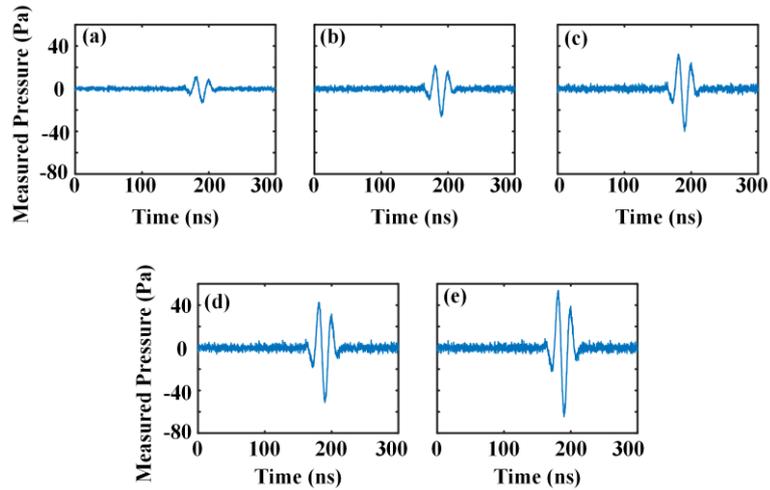

Figure 3. Superposition of photoacoustic signals.

The five photoacoustic signals are superimposed on each other to obtain one superimposed photoacoustic signals, two superimposed photoacoustic signals, three superimposed photoacoustic signals, four superimposed photoacoustic signals, and five superimposed photoacoustic signals, respectively, as shown in Figures 3(a)-3(e). As the number of superimposed laser pulses increases, the intensity of the photoacoustic signal and the average level of noise are also increasing. However, the former increases far beyond the latter, and the signal-to-noise ratio of the signal increases with the increase of the number of laser pulses, as shown in Figure 4. As the number of laser pulses increase from one to five, the signal-to-noise ratio is 24.95 dB, 27.36 dB, 29.01 dB, 30.10 dB and 31.30 dB, respectively. The experimental results can prove that in A-scan mode, it is completely feasible to increase the signal-to-noise ratio by superimposing multiple laser pulses.

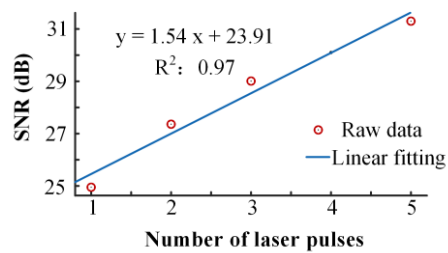

Figure 4. The relationship between the signal-to-noise ratio and the number of superimposed laser pulses.

### 3.2 B-Scan: Multi-pulse imaging

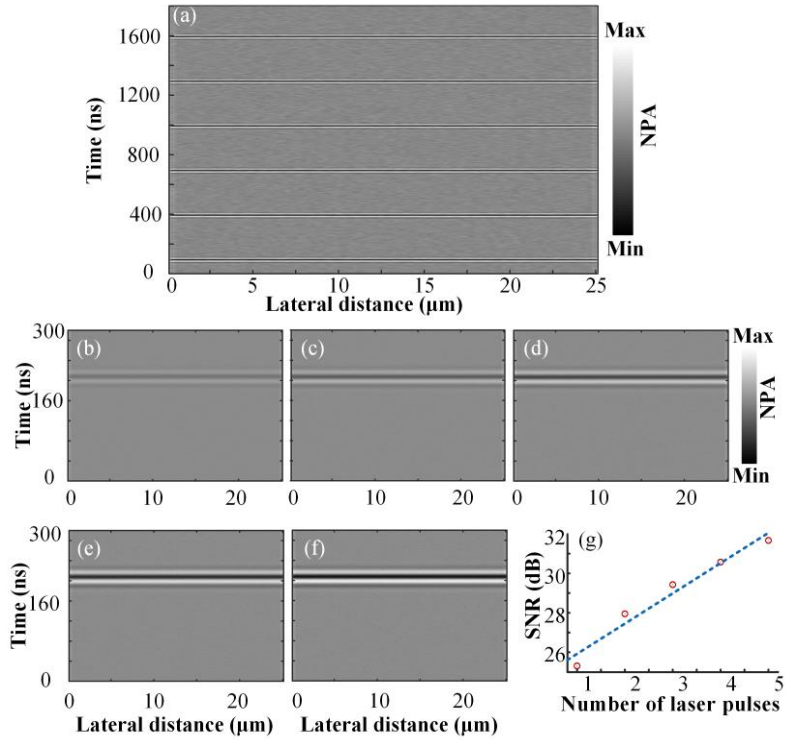

Figure 5. B-scan. (a) B-scan image of Carbon fiber using 6 laser pulses. (b)-(f) are one superimposed B-scan image, two superimposed B-scan image, three superimposed B-scan image, four superimposed B-scan image, and five superimposed B-scan image, respectively. (g) signal-to-noise ratio vs. the number of superimposed laser pulses. NPA, normalized photoacoustic amplitude.

The experimental results of the A-scan mode confirmed the feasibility of this method, and the B-scan mode was further used to verify the feasibility of the multi-pulse system. Carbon fiber is used for imaging to obtain a two-dimensional B-scan image, as shown in Figure 5(a). The brighter part in the figure is the photoacoustic signal, and in the depth direction, the time interval between the two photoacoustic signals is 300 ns; The darker part is a noise signal, and its intensity is much lower than the intensity of the photoacoustic signal. Figures 5(b)-5(f) are one superimposed B-scan image, two superimposed B-scan image, three superimposed B-scan image, four superimposed B-scan image, and five superimposed B-scan image, respectively. And the signal-to-noise ratios for Figures 5(b)-5(f) are 25.31 dB, 27.95 dB, 29.43 dB, 30.57 dB and 31.67 dB, respectively, as shown in Figure 5(g). The signal-to-noise ratio of the signal increases linearly with the number of superimposed pulses.

**3.3 C-Scan**

A binary vascular network is used to further demonstrate the effectiveness of the method, as shown in Figure 6(a). Figure 6(b) is the MAP image, due to the presence of noise, the image looks a little fuzzy and the signal-to-noise ratio is not high. Figures 6(c)-6(g) are one superimposed MAP image, two superimposed MAP image, three superimposed MAP image, four superimposed MAP image, and five superimposed MAP image, respectively. When only one laser pulse is used for imaging, the signal-to-noise ratio of the image is not high (~ 25.93 dB), and the noise is obvious, as shown in Figure 6(b). When two laser pulses are used for each A-scan, the image becomes smoother, and the signal-to-noise ratio is higher (~ 28.45 dB). Similarly, when three laser pulses are used to irradiate the sample sequentially, the signal-to-noise ratio is further improved (~ 29.96 dB), and the blood vessels in the image become clearer. Continue to increase the number of laser pulses, and the signal-to-noise ratio becomes higher. For example, the signal-to-noise ratio is 31.13 dB and 32.21 dB when four and five laser pulses are used, respectively.

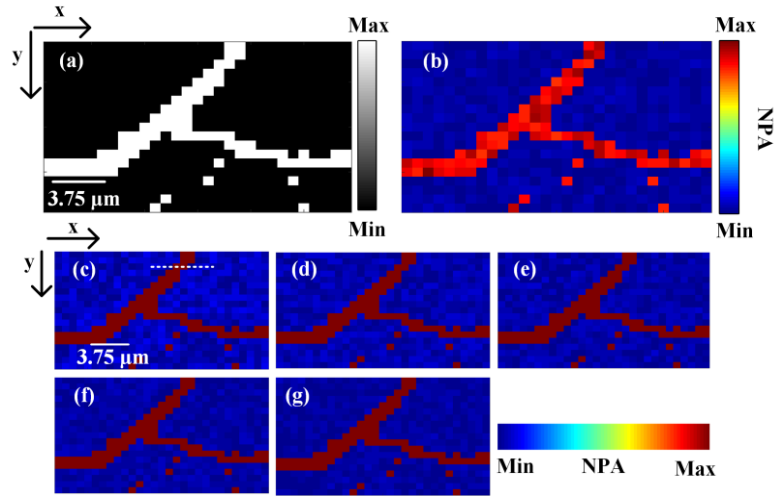

Figure 6. C-scan. (a) A binary vascular network. (b) MAP image of vascular network using 6 laser pulses. (b)-(f) are one superimposed MAP image image, two superimposed MAP image image, three superimposed MAP image image, four superimposed MAP image image, and five superimposed MAP image image, respectively.

**4 Conclusion**

In summary, we have developed a multi-focus photoacoustic microscopy. In each A-line data acquisition, multiple laser pulses are used to sequentially irradiate the sample to generate photoacoustic signal, and then the photoacoustic signals are superimposed on each other to achieve promotion of photoacoustic image quality. In order to verify the feasibility of this method, a multi-pulse photoacoustic imaging simulation platform was established using the k-Wave toolbox. The performance of multi-pulse photoacoustic imaging is verified by the three scanning modes of photoacoustic microscope, A-scan, B-scan and C-scan. The results show that SNR is proportional to the number of laser pulses used. The higher signal-to-noise ratio wih lower laser pulse energy will make the photoacoustic microscope more widely used in the biomedical field.